\documentclass[reprint,aps,prb,superscriptaddress]{revtex4-2}

\usepackage{graphicx}
\usepackage{comment}
\usepackage{hyperref}
\usepackage{amsmath,amssymb,amsfonts}
\usepackage{color}
\usepackage{xcolor}
\usepackage{subcaption}
\usepackage{caption}
\usepackage{orcidlink}
\hypersetup{colorlinks=true}

\begin{document}


\preprint{APS/123-QED}

\title{Interactions of composite magnetic skyrmion-superconducting vortex pairs in ferromagnetic superconductors}

\author{Paul~Leask\orcidlink{0000-0002-6012-0034}}\email{palea@kth.se}
\affiliation{Department of Physics, KTH Royal Institute of Technology, 10691 Stockholm, Sweden}
\author{Calum~Ross\orcidlink{0000-0001-6728-1239}}
\affiliation{Department of Computer Science, Edge Hill University, St Helens Rd., Ormskirk L39 4QP, UK}
\affiliation{Research and Education Center for Natural Sciences, Keio University, Hiyoshi 4-1-1, Yokohama, Kanagawa 223-8521, Japan}
\author{Egor~Babaev\orcidlink{0000-0001-7593-4543}}
\affiliation{Department of Physics, KTH Royal Institute of Technology, 10691 Stockholm, Sweden}
\affiliation{Wallenberg Initiative Materials Science for Sustainability, Department of Physics, KTH Royal Institute of Technology, 10691 Stockholm, Sweden}

\date{\today}

\begin{abstract}
We study composite topological excitations in ferromagnetic superconductors consisting of bound states of magnetic spin textures (skyrmions) and superconducting vortices.
Using a Ginzburg--Landau framework with Zeeman coupling between the magnetization and superconducting magnetic field, we demonstrate that skyrmion-vortex pairs (SVPs) form energetically stable bound states.
By analyzing their asymptotic interactions, we identify regimes in which SVPs exhibit both short-range repulsion and long-range attraction, leading to clustering phenomena.
Our results provide a field-theoretical basis for understanding suggest pathways for controlling hybrid topological matter through long-range interactions.
\end{abstract}

\maketitle



\section{Introduction}
\label{sec: Introduction}

The possibility of superconductivity occurring in a ferromagnetic material was first addressed by Ginzburg \cite{Ginzburg_1957}.
Models for coexisting magnetic and superconducting states were later proposed \cite{Tachiki_1979,Varma_1979,Varma_1981} by combining Ginzburg--Landau theory with a mean field theoretic model of the magnetic subsystem.
This enabled the properties of ferromagnetic superconductors to be modeled.
In such systems, the magnetic order arises from local moments, while superconductivity is carried by conduction electrons.
In such systems, the magnetic order parameter can coexisting with superconductivity is carried by conduction electrons in a range of temperatures $T_m<T_C$.
For ferromagnetic superconductors, there exists a stable temperature range below $T_m<T_C$, in which the magnetization vector has fixed length, as this minimizes the free energy.
This existence of two order parameters allows topological magnetic spin textures that coexist with superconducting states.

A particularly interesting example of such a composite topological excitation is the magnetic skyrmion-superconducting vortex pair (SVP), that were suggested in spatially separated superconductor-ferromagnet bilayers \cite{Andriyakhina_2021,Pershoguba_2016,Dahir_2019,Genkin_1994,Derendorf_2024,Hals_2016,Menezes_2019,Valle_2015} and claimed to be observed experimentally in chiral magnet–superconductor heterostructures \cite{Petrovi_2021,Xie_2024}.
In addition, SVPs have recently been studied in the bulk of ferromagnetic superconductors, rather than on the interface of a magnet-superconductor heterostructure \cite{Mukherjee_2025}.
Moreover, SVPs  have been proposed as platforms for Majorana bound states \cite{Rex_2019}, where the bound zero modes can be manipulated, braided \cite{Nothhelfer_2022}, or form chiral Majorana bands \cite{Garnier_2019}.
Moreover, SVPs  have been proposed as platforms for Majorana  states \cite{Rex_2019, Nothhelfer_2022,Garnier_2019}.
These proposals further motivate an understanding of the energetics and orientation-dependent interactions of SVPs.
We are interested in such interactions of SVPs in the bulk of ferromagnetic superconductors.

In magnet-superconductor heterostructures, superconducting vortices are usually approximated by thin-film Pearl vortices \cite{Pearl_1964}, with no back-reaction on the superconducting order parameter \cite{Apostoloff_2023,Apostoloff_2024}.
Under this assumption, the superconductor merely provides an external, spatially varying magnetic field that acts on the underlying magnet, effectively reducing the problem to that of a magnetic system subjected to an \emph{inhomogeneous} applied field.
While this simplification has yielded valuable insights (such as skyrmion deformation, chirality inversion, and field-induced stabilization), it neglects the mutual feedback between the superconducting and magnetic subsystems.
In reality, the screening currents and the magnetization texture influence one another, modifying both the local field distribution and the energy landscape.
Likewise, the full back-reaction in the bulk of ferromagnetic superconductors has also not been accounted for, in order to obtain a tractable equation for the magnetic spin texture \cite{Mukherjee_2025}.
Accounting for this back-reaction is crucial to capture the full range of possible bound states and interaction forces between composite excitations.
Our aim is to do this self-consistently in the bulk of a ferromagnetic superconductor, properly accounting for the back-reaction on the superconductor.
 
The theoretical and experimental pursuit motivates an understanding of the energetics and orientation-dependent interactions of SVPs.
Interactions among topological solitons have been extensively studied across a variety of field theories.
In superconductors, the Abelian Higgs model captures the essential physics of vortex-vortex forces, where exponentially decaying scalar and gauge field modes compete to determine whether the interaction is attractive or repulsive \cite{Winyard_2025,Speight_1997,Manton_Speight_2003}, and similar structures appear in cosmic string models \cite{Bettencourt_1995,Fujikura_2023}.
Analogous analysis of long-range intervortex forces have been conducted in imbalanced fermionic systems \cite{Barkman_2020}, competing-order superconductors \cite{Speight_2021} and in Chern–Simons–Landau–Ginzburg theories of the fractional quantum Hall effect \cite{leask2025anyonboundstateshybrid}.
Interactions among skyrmions have likewise been studied in detail, from chiral ferromagnetic systems with Dzyaloshinskii–Moriya interactions \cite{Schroers_2023}, to skyrmions in nuclear matter \cite{Manton_2004} and analogous baby skyrmion models \cite{Piette_1995,Speight_2010}.
However, the case of composite skyrmion-vortex pairs, where magnetic and superconducting textures coexist and {\it interact self-consistently}, remains not investigated.

In this paper, we address this gap by studying a self-consistent model of a ferromagnetic superconductor, where both superconducting and magnetic fields are treated on equal footing with the associated back-reaction included. 
This framework allows us to construct and analyze SVPs using a combination of analytic and numerical methods.
In particular, we focus on the far-field asymptotics and long-range interactions of SVPs, identifying the conditions under which competing attractive and repulsive forces lead to the formation of bound states within this coupled system.


\section{The superconducting ferromagnetic model}
\label{sec: The superconducting ferromagnetic model}

The model we are interested in is that of an isotropic ferromagnetic superconductor proposed in \cite{Varma_1979,Varma_1981}.
It consists of a superconducting order parameter which is a single complex field $\psi\in\mathbb{C}$, where $|\psi|^2$ is a measure of local density of Cooper pairs, an electromagnetic gauge field $\vec{A}=(A_x,A_y,A_z)\in\mathbb{R}^3$, and a magnetization order parameter $\vec{m}=(m_x,m_y,m_z)\in \mathbb{R}^3$.
We are interested in translation invariant solutions, with the translation invariance imposed in the $z$-direction.
Then,  associated to the gauge field is the magnetic field
\begin{equation}
    \vec{B}=\vec{\nabla}\times\vec{A}=(\partial_y A_z,-\partial_x A_z,\partial_x A_y-\partial_y A_x).
\end{equation}
The free energy functional of this system consists of three parts
\begin{equation}
    F[\psi,\vec{A},\vec{m}] = F_{\textup{sc}}[\psi,\vec{A}] + F_{\textup{mag}}[\vec{m}] + F_{\textup{int}}[\psi,\vec{A},\vec{m}].
\end{equation}

The first part is the free energy functional for the superconductor $(\psi,\vec{A})$, which is given by the Ginzburg--Landau free energy density
\begin{equation}
    \begin{split}
        \mathcal{F}_{\textup{sc}}[\psi,\vec{A}] = \frac{1}{2}|\vec{D}\psi|^2 + \frac{1}{2}|\vec{B}|^2  + \frac{a(T)}{2}|\psi|^2 + \frac{b}{4}|\psi|^4
    \end{split},
\end{equation}
where $\vec{D}\psi=\vec{\nabla}\psi + iq\vec{A}\psi$ is the gauge covariant derivative and $a(T)=a_0(T-T_c)/T_c$.
The critical temperature for superconductivity is $T_c$ and $q\sim2e$ is the effective charge of a Cooper pair.
For the magnetization we consider an isotropic ferromagnet in the absence of an applied magnetic field.
In the simplest approximation, the free energy is given by
\begin{equation}
    \mathcal{F}_{\textup{mag}}[\vec{m}] =  \frac{\alpha(T)}{2}|\vec{m}|^2 + \frac{\beta}{4}|\vec{m}|^4 + \frac{1}{2}|\nabla\vec{m}|^2,
\end{equation}
where $\alpha(T)=\alpha_0(T-T_m)/T_m$ and $T_m$ is the Curie temperature.
Throughout we will consider materials in which they exhibit superconductivity above their Curie temperature, $T_c > T_m$ \cite{Liu_2016,Nandi_2014,Devizorova_2019}.
We are interested in superconducting vortices in the presence of magnetic spin textures such as skyrmions.
So, in this work, we consider a regime where  we can restrict the magnetization to be of fixed length, that is $\vec{m}\in \mathbb{S}^2_{m_0} \subset\mathbb{R}^3$.

There are two main interactions of the superconducting state $(\psi,\vec{A})$ with the magnetization $\vec{m}$.
One is via the direct effects of spin-flip scattering of conduction electrons with the magnetic moments and conduction-electron polarization.
The second is an indirect interaction which arises from the coupling of the order parameter $\psi$ to the electromagnetic gauge field $\vec{A}$, and the coupling of the magnetic field $\vec{B}=\vec{\nabla}\times\vec{A}$ to the magnetization $\vec{m}$ through the Zeeman interaction \cite{Batista_2012}
\begin{equation}
    \mathcal{F}_{\textup{Zeeman}}[\vec{A},\vec{m}] = - \vec{m} \cdot (\vec{\nabla}\times\vec{A}).
\label{eq: Zeeman interaction}
\end{equation}
We first begin by ignoring the effects of spin-flip scattering and consider the interaction energy functional defined by
\begin{equation}
\label{eq: Free energy SC}
    F_{\textup{int}}[\psi,\vec{A},\vec{m}] = F_{\textup{Zeeman}}[\vec{A},\vec{m}].
\end{equation}

The uniform ground state configurations for the superconducting order parameter $\psi$ and the magnetization $\vec{m}$ are determined by minimizing the potential energy
\begin{equation}
    \mathcal{F}_p=\frac{a}{2}|\psi|^2 + \frac{b}{4}|\psi|^4 + \frac{\alpha}{2}|\vec{m}|^2 + \frac{\beta}{4}|\vec{m}|^4.
\end{equation}
which amounts to solving the system of equations
\begin{align}
    \left.\frac{\delta \mathcal{F}_p}{\delta |\psi|}\right|_{(u,m_0)} = \, & au + bu^3 = 0, \\ \left.\frac{\delta \mathcal{F}_p}{\delta |\vec{m}|}\right|_{(u,m_0)} = \, & \alpha m_0 + \beta m_0^3  = 0.
\end{align}
This gives us the ground state configurations
\begin{equation}
    u^2 = -\frac{a}{b}, \quad m_0^2 = -\frac{\alpha}{\beta}.
\end{equation}
The corresponding ground state free energy density is determined to be 
\begin{equation}
    \mathcal{F}_p^* = -\frac{a^2}{4b} - \frac{\alpha^2}{4\beta}.
    \label{eq: gs free energy}
\end{equation}
Using this free energy we can understand the phases of the model.


\section{Composite magnetic skyrmion-superconducting vortex pair}
\label{sec: Composite magnetic skyrmion-superconducting vortex pair}

In order to study the interactions of composite SVPs, it will prove convenient to normalize the energy such that the ground state configuration has zero energy.
To do this, we consider the non-linear sigma model limit by requiring the magnetization to have fixed length $|\vec{m}|=m_0$.
That is, we define the normalized free energy of the theory to be
\begin{align}
\label{eq: Normalized SC energy}
    E = \, & \int_{\mathbb{R}^2} \textup{d}^2x \left\{ \frac{1}{2}|\vec{D}\psi|^2 + \frac{1}{2}|\vec{\nabla}\times\vec{A}|^2 + \frac{b}{4} \left( u^2 - |\psi|^2 \right)^2 \right. \nonumber \\
    \, & \left. + \frac{1}{2}|\nabla\vec{m}|^2 - \vec{m}\cdot(\vec{\nabla}\times\vec{A})\right\}.
\end{align}

In this model, we are interested in stationary configurations that take the form of local minima of the free energy \eqref{eq: Normalized SC energy}.
These satisfy the (bulk) ferromagnetic Ginzburg--Landau equations that are obtained by variation of $E$ with respect to the fields $(\psi,\vec{A},\vec{m})$, which yields the Euler-Lagrange field equations
\begin{subequations}
\label{eq: SC EL}
    \begin{align}
        \label{eq: SC EL - OP}
        \frac{\delta E}{\delta \psi^*} = \, & -b\psi \left( u^2 - |\psi|^2 \right) - \frac{1}{2}\vec{D}\cdot\vec{D}\psi = 0, \\
        \label{eq: SC EL - Gauge}
        \frac{\delta E}{\delta \vec{A}} = \, & q^2|\psi|^2 \vec{A} + \frac{iq}{2}\left( \psi \vec{\nabla}\psi^* - \psi^*\vec{\nabla}\psi \right) + \vec{\nabla}\times\vec{\nabla}\times\vec{A} \nonumber \\
        \, & - \vec{\nabla}\times\vec{m} = \vec{0},\\
        \label{eq: SC EL - Magnet}
        \frac{\delta E}{\delta \vec{m}} = \, & - \Delta \vec{m}  - \vec{\nabla}\times\vec{A} = \vec{0}.
    \end{align}
\end{subequations}
From the gauge field equation \eqref{eq: SC EL - Gauge}, we get the supercurrent
\begin{align}
    \vec{J} = \frac{iq}{2}\left( \psi \vec{\nabla}\psi^* - \psi^*\vec{\nabla}\psi \right) + q^2|\psi|^2 \vec{A},
\end{align}
and the magnetization current
\begin{equation}
    \vec{J}_{\textup{mag}} = \vec{\nabla}\times\vec{m}.
\end{equation}

We now detail our numerical methods, that are implemented for NVIDIA CUDA architecture, for finding topological solitons in this model of ferromagnetic superconductors.
These are field configurations $\{\psi,\vec{A},\vec{m}\}$ that minimize the energy functional \eqref{eq: Normalized SC energy} and are constrained by the topology of the manifold of degenerate vacuum states.
In particular, we are interested in obtaining magnetic skyrmion solutions in the magnetization $\vec{m}$ and vortices in the superconductor $\{\psi,\vec{A}\}$.
These are solutions to the Euler--Lagrange field equations \eqref{eq: SC EL} and have some non-trivial topological invariants associated to them.

To obtain multi magnetic skyrmion and superconducting vortex states, we choose to numerically relax the energy \eqref{eq: Normalized SC energy} using arrested Newton flow - an accelerated second order gradient descent based method with flow arresting criteria.
For numerical purposes, it is convenient to concatenate our fields into one unified field $\phi=(\psi,\vec{A},\vec{m})\in\mathbb{C}\times\mathbb{R}^3\times S^2 \subset\mathbb{R}^8$ and employ a fourth order central finite difference method.
This is performed on a meshgrid of $N^2$ lattice sites with lattice spacing $h$ and results in a discrete approximation $E_{\textup{dis}}(\phi)$ to the energy $E(\psi,\vec{A},\vec{m})$.
Then the discretised energy can be thought of as a function $E_{\textup{dis}}:\mathbb{R}^{8N^2} \rightarrow \mathbb{R}$.

The algorithm is conceptually rather simple.
The basis of the arrested Newton flow is detailed as follows: starting from rest $\dot{\phi}(0)=\vec{0}$, we solve for the motion of a particle in the discretised configuration space $\mathcal{C}$ under the potential $E_{\textup{dis}}$, 
\begin{equation}
    \ddot{\phi} = -\textup{grad} E_{\textup{dis}}(\phi),
\label{eq: Arrested Newton flow}
\end{equation}
with some appropriate initial configuration $\phi(0)$.
In terms of the order parameters, this is reformulating the minimization as a second order dynamical problem and solving the second order coupled system
\begin{align}
    \frac{\textup{d}^2\psi}{\textup{d}t^2} = \, & \frac{1}{2}D_i D_i \psi +b\psi\left( u^2-|\psi|^2 \right), \\
    \frac{\textup{d}^2 A_i}{\textup{d}t^2} = \, & \partial_j (\partial_j A_i - \partial_i A_j) - J_i + \epsilon_{ijk}\partial_j m_k, \\
    \frac{\textup{d}^2 m_i}{\textup{d}t^2} = \, & \partial_j \partial_j m_i + \epsilon_{ijk} \partial_j A_k,
\end{align}
where $t$ is a fictitious time coordinate.
We can reduce this problem to a coupled first order system and we solve the coupled system simultaneously with a fourth order Runge--Kutta method.
After each time step $t \mapsto t + \delta t$, we compare energies of the current and previous configurations.
If the energy has increased, $E_{\textup{dis}}(t + \delta t) > E_{\textup{dis}}(t)$, we arrest the flow by setting $\dot{\phi}(t + \delta t)=\vec{0}$ and restart it from the current configuration.
The flow is deemed to have converged when $\lVert E_{\textup{dis}}(\phi) \rVert_\infty<\epsilon$, where $\epsilon$ is some threshold tolerance (we use $\epsilon=10^{-6}$).

\begin{figure*}[t]
    \centering
    \includegraphics[width=0.8\linewidth]{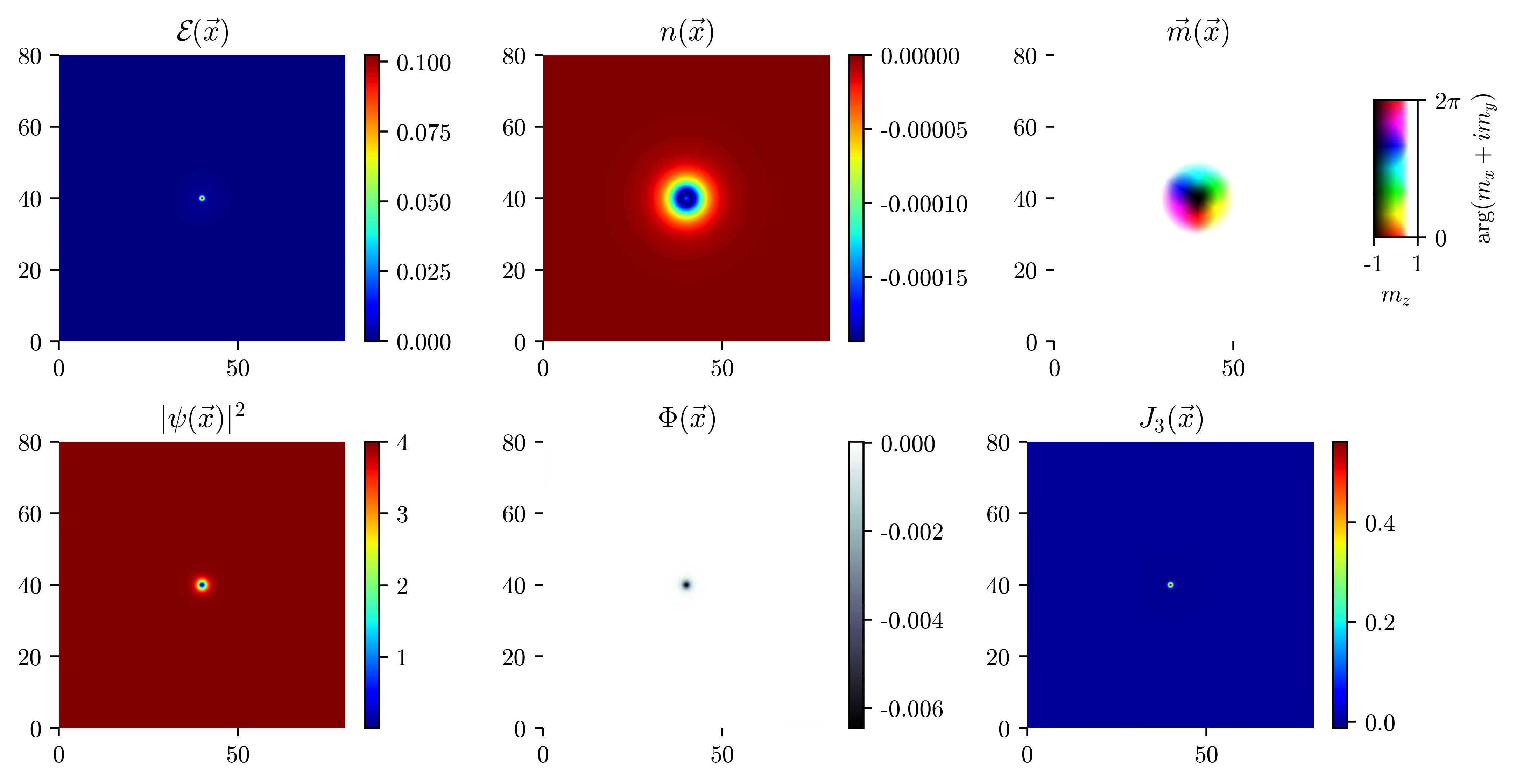}
    \caption{A composite skyrmion-vortex pair, consisting of a $N=1$ superconducting vortex and a $n=-1$ magnetic Bloch skyrmion in the mixed superconducting/ferromagnetic phase. They are obtained by numerically solving the full field theory equations of motions \eqref{eq: SC EL}, associated to the normalized energy \eqref{eq: Normalized SC energy}, with the arrested Newton flow minimization method. The parameters used here are $a=-1, b=1/4$ and $\beta=-\alpha=1$, which gives the vacuum values $m_0^2=1$ and $u^2=4$. The associated decay length for the superconducting order parameter is $\xi_s=1/\sqrt{2}$, the magnetization decay length is $ l_m =1/\sqrt{3}$ and the magnetic penetration depth is $\lambda=1/2$.}
\label{fig: N=1,Q=1 SC}
\end{figure*}

For the superconducting order parameter we use an extended version of the Nielsen--Olesen ansatz \cite{Nielsen_1973,Hindmarsh_1995}
\begin{equation}
\label{eq: Nielsen-Olesen ansatz}
    \begin{split}
        \psi(r,\theta) = \,& \sigma(r)e^{-iN\theta}, \\
        \vec{A}(r,\theta) = \, & \left(-\frac{a(r)}{r}\sin\theta,\frac{a(r)}{r}\cos\theta,g(r)\right),
    \end{split}
\end{equation}
where the profile functions satisfy the boundary conditions $\sigma(0) = 0, \sigma(\infty) = u$, $a(0)=0, a(\infty)=N/q$ and $g'(0)=g(\infty)=0$, and $N\in\mathbb{Z}$ is the winding number, or vortex number.
Now, by Stoke's theorem, it follows that the total magnetic flux through the $xy$-plane is thus
\begin{equation}
    \Phi = \int_{\mathbb{R}^2} B_3 \textup{d}^2x = 2\pi \int_0^\infty \frac{\textup{d}a}{\textup{d}r} \textup{d}r = N\frac{2\pi}{q}\equiv N\Phi_0,
\end{equation}
and
\begin{equation}
    \int_{\mathbb{R}^2} B_1 \textup{d}^2x = \int_{\mathbb{R}^2} B_2 \textup{d}^2x = 0.
\end{equation}
Hence, the flux of the vortices are quantized, with $\Phi_0=2\pi/q$ being the quantum. 
For the magnetization we use the Bloch ansatz \cite{Bogdanov_1994}
\begin{equation}
\label{eq: Bloch ansatz}
    \vec{m}(r,\theta)=
    \begin{pmatrix}
        -\sqrt{1-f(r)^2} \sin(Q\theta) \\
        \sqrt{1-f(r)^2} \cos(Q\theta) \\
        f(r)
    \end{pmatrix},
\end{equation}
where $f(r)$ is some monotonically increasing profile function that satisfies the boundary conditions $f(0)=-1$ and $f(\infty)=1$.
It can be seen that at $r=0$ we have spin down states whereas we have spin up states as $r\rightarrow\infty$.
There is also a topological invariant associated with the magnetization ansatz, which is
\begin{align}
    n =  \frac{1}{4\pi} \int_{\mathbb{R}^2}\vec{m} \cdot \left( \partial_x \vec{m} \times \partial_y \vec{m} \right)  \textup{d}^2x = -Q \in \mathbb{Z}.
\end{align}

We note that there are other ans\"atze that we could have employed for the magnetization, such as the antiskyrmion and N\'eel hedgehog skyrmion considered in \cite{Leask_speight_2026}, but these yield skyrmions with higher energy than Bloch skyrmions in this model.
Hence, throughout we restrict our analysis to SVPs of the Bloch type.

The composite magnetic skyrmion-superconducting vortex pair, consisting of a $N=1$ superconducting vortex and a $n=-1$ magnetic Bloch skyrmion, can be obtained by using the Nielsen--Olesen \eqref{eq: Nielsen-Olesen ansatz} and Bloch \eqref{eq: Bloch ansatz} ans\"atze as initial configurations for the arrested Newton flow algorithm.
The numerically relaxed axially symmetric SVP is a solution of the static field equations \eqref{eq: SC EL} and is plotted in Fig.~\ref{fig: N=1,Q=1 SC}.

\begin{figure*}[t]
    \centering
    \includegraphics[width=0.8\linewidth]{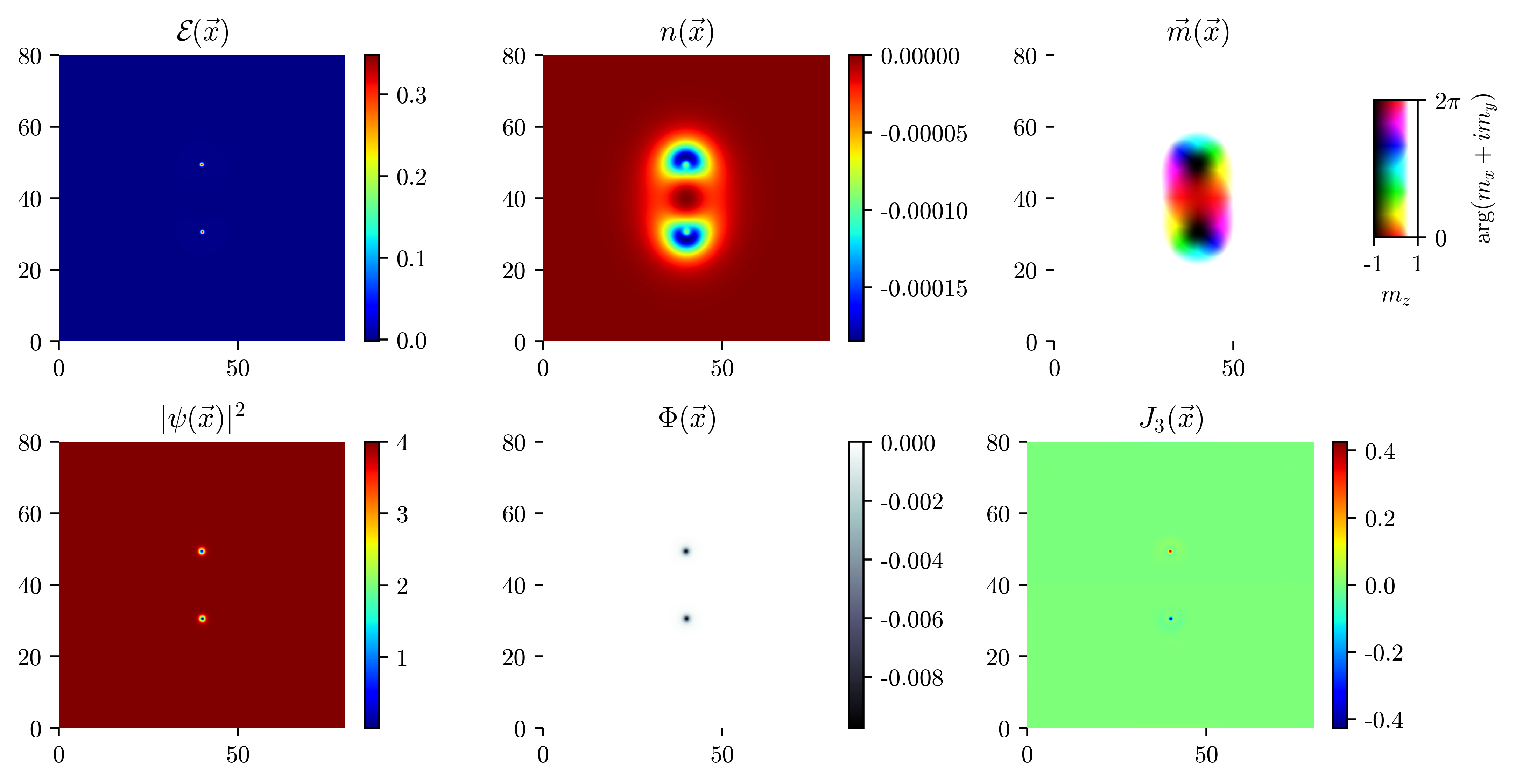}
    \caption{A bound state of composite SVPs, where each SVP experiences short-range repulsion and long-range attraction. The parameters used here are $a=-4, b=1$ and $\beta=-\alpha=1$, which gives the vacuum values $m_0^2=1$ and $u^2=4$. The associated decay length for the superconducting order parameter is $\xi_s=1/\sqrt{8}$, the magnetization decay length is $ l_m =1/\sqrt{3}$ and the magnetic penetration depth is $\lambda=1/2$. Therefore we are in the bound state regime since $\xi_s<\lambda< l_m $ and the binding energy is negative, $E_{\textup{SVP-SVP}}-2E_{\textup{SVP}}<0$.}
\label{fig: N=2,Q=2 SC}
\end{figure*}

To construct an initial configuration for multiple SVPs, we can use two different methods.
The first is straightforward where we consider axially symmetric initial configurations and simply set $N>1$ and $Q>1$.
While these ans\"atze are initially axially symmetric, they do not necessarily relax to axially symmetric configurations.
The second method involves separated SVPs, where the soliton cores do not overlap.
For the magnetization, this is carried out using the $\mathbb{C}P^1$ formalism.
That is, we introduce the complex variable
\begin{equation}
\label{eq: CP1 formalism}
    W(\vec{x}) = \frac{m_x(\vec{x})+im_y(\vec{x})}{1+m_z(\vec{x})} \in \mathbb{C}.
\end{equation}
Then we can construct multi-skyrmions, of topological degree $n=-k$, using the product ansatz
\begin{equation}
\label{eq: Product ansatz}
    W(\vec{x}) = \sum_{i=1}^{k} W_i(\vec{x}-\vec{x}_i),
\end{equation}
where $\vec{x}_i$ is the location of the $i$th SVP and we use the Bloch ansatz \eqref{eq: Bloch ansatz} with $Q=1$ for each $W_i$.
The resulting magnetization can be recovered via
\begin{equation}
    \vec{m} = \frac{1}{1+|W|^2}
    \begin{pmatrix}
        W+W^* \\
        i(W^*-W) \\
        1-|W|^2
    \end{pmatrix}.
\end{equation}
In a similar manner, we introduce the Abrikosov ansatz \cite{Abrikosov_1957} to obtain a superconducting $k$-vortex,
\begin{equation}
\label{eq: Abrikosov ansatz}
    \psi(\vec{x}) = \prod_{i=1}^{k} \psi_i(\vec{x}-\vec{x}_i), \quad \vec{A}(\vec{x}) = \sum_{i=1}^{k} \vec{A}_i(\vec{x}-\vec{x}_i).
\end{equation}
Together, these ans\"atze construct $k$-separated SVPs.

Let us focus on the case of interest, $k=2$, which describes two initially separated SVPs, $\vec{m}_1(\vec{x}_1)$ and $\vec{m}_2(\vec{x}_2)$.
Motivated by the interactions of skyrmions in the baby Skyrme model, we allow for a relative (iso-)rotation between the skyrmions.
That is, we rotate one of the skyrmions, say $\vec{m}_2(\vec{x}_2)$, by the mapping $\vec{m}_2(\vec{x}_2)\mapsto R(\chi)\vec{m}_2(\vec{x}_2)$, where
\begin{equation}
    R(\chi)=
    \begin{pmatrix}
        \cos\chi & -\sin\chi & 0\\
        \sin\chi & \cos\chi & 0 \\
        0 & 0& 1
    \end{pmatrix}
    \in\textup{SO}(3).
\end{equation}
Then we construct the separated SVPs using the $\mathbb{C}P^1$ formalism \eqref{eq: CP1 formalism}, along with the product ansatz \eqref{eq: Product ansatz}, for the skyrmion and the Abrikosov ansatz \eqref{eq: Abrikosov ansatz} for the vortex.
For a particular choice of parameters, the formation of a stable SVP-SVP bound state is observed.
This SVP-SVP bound state is shown in Fig.~\ref{fig: N=2,Q=2 SC}.
It is found that the force between the separated SVPs is repulsive for $\chi=0$ and most attractive for $\chi=\pi$.
This is akin to what happens in the baby Skyrme model \cite{Piette_1995}.
Skyrmions orientated with $\chi=\pi$ are said to be in the attractive channel orientation, whereas for $\chi=0$ they are in the repulsive channel.
We now look to understand the interactions between SVPs and explain the mechanism behind the formation of stable SVP-SVP bound states.


\section{Long-range interactions of composite skyrmion-vortex pairs}
\label{sec: Long-range interactions of composite SVPs}


\subsection{Asymptotic form of the composite state}
\label{subsec: Asymptotic form of the composite state}

The free energy density we consider is given by
\begin{equation}
\label{eq: Normalized free energy}
    \begin{split}
        \mathcal{F} = \frac{1}{2}\overline{D_i\psi}D_i\psi + \frac{1}{4}F_{ij}F_{ij} + \frac{b}{4}\left( u^2-|\psi|^2 \right)^2 \\+ \frac{1}{2}\partial_j m_i \partial_j m_i - \epsilon_{ijk}m_i \partial_jA_k
    \end{split},
\end{equation}
where $F_{ij}=\partial_i A_j - \partial_j A_i$ is the field strength and we have normalized the energy by subtracting off the ground state energy.
Let us linearize about the ground state $\{u,\vec{0},\vec{m}_0\}$ in ferromagnetic superconducting phase,
\begin{equation}
\label{eq: Linearized variables}
    \psi = u + \phi, \quad \vec{A} = \vec{0} + \vec{\alpha}, \quad \vec{m} = \vec{m}_0 + \vec{n}.
\end{equation}
Without loss of generality, we set the fixed length of the magnetization to be unity, i.e. $|\vec{m}|^2=1$.
To determine the form of the magnetization perturbation $\vec{n}$, we must consider the expansion \cite{Piette_1995}
\begin{equation}
    \vec{m} = \sqrt{1-\vec{n}\cdot\vec{n}} \, \vec{m}_0 + \vec{n} \approx \vec{m}_0 + \vec{n} + O(\vec{n}\cdot\vec{n}).
\end{equation}
Taking the magnitude of this gives us the constraint
\begin{align}
    \vec{m}\cdot\vec{m} = \, & 1 + 2\sqrt{1-\vec{n}\cdot\vec{n}} \, (\vec{m}_0\cdot \vec{n}) \overset{!}{=} 1.
\end{align}
Therefore, we see that the magnetization perturbation must be orthogonal to the ground state magnetization, that is
\begin{equation}
    \vec{m}_0\cdot\vec{n}=0 \quad \Rightarrow \quad \vec{n}\in T_{\vec{m}_0}S^2.
\end{equation}
This means that the only low-energy excitations are transverse (directional) modes with an associated decay length, i.e. there is no longitudinal amplitude mode nor associated coherence length in this type of models \cite{Kosevich_1990}.

The linearized free energy is obtained by substituting the linearized fields \eqref{eq: Linearized variables} into the free energy \eqref{eq: Normalized free energy} and considering terms up to quadratic order in the fields.
The resulting linearized energy is
\begin{equation}
    \begin{split}
        \mathcal{F}_{\textup{lin}} = \frac{1}{2}|\vec{\nabla}\phi|^2 + bu^2\phi^2 + \frac{1}{2}|\vec{\nabla}\times\vec{\alpha}|^2 + \frac{1}{2}q^2u^2|\vec{\alpha}|^2 \\+ \frac{1}{2}|\nabla\vec{n}|^2 - \vec{n}\cdot(\vec{\nabla}\times\vec{\alpha})
    \end{split}.
\end{equation}
To determine the asymptotic form of the fields, we need to compute the variation of the linearized energy with respect to the perturbations $\{\phi,\vec{\alpha},\vec{n}\}$.
We find that the superconducting order parameter $\phi$ is described by a \emph{Klein-Gordon} equation
\begin{equation}
    \frac{\delta \mathcal{F}_{\textup{lin}}}{\delta \phi} = \left( -\Delta + 2bu^2 \right) \phi = 0.
\end{equation}
The electromagnetic gauge field $\vec{\alpha}$ satisfies a \emph{Proca} equation, with source generated by the (curl of the) magnetization,
\begin{equation}
    \frac{\delta \mathcal{F}_{\textup{lin}}}{\delta \vec{\alpha}} = -\Delta\vec{\alpha} + \vec{\nabla}(\vec{\nabla}\cdot\vec{\alpha}) + q^2u^2\vec{\alpha} - \vec{\nabla}\times\vec{n} = \vec{0}.
\end{equation}
Finally, the magnetization $\vec{n}$ is determined by a vector \emph{Poisson} equation, where the magnetic field provides the source,
\begin{equation}
    \frac{\delta \mathcal{F}_{\textup{lin}}}{\delta \vec{n}} = -\Delta \vec{n} - \vec{\nabla}\times\vec{\alpha} = \vec{0}.
\end{equation}

As the isolated single skyrmion-vortex pair is axially symmetric, we consider field perturbations that also have axial symmetry.
Then, the linearized field equation for superconducting order parameter $\phi(r)$ reduces to Bessel's modified equation of zeroth order,
\begin{equation}
    r^2\frac{\textup{d}^2\phi}{\textup{d} r^2} + r\frac{\textup{d}\phi}{\textup{d} r} - 2bu^2r^2\phi = 0.
\end{equation}
The solutions to this are well-known and are given by $\phi(r) = c_{\psi} K_0\left(\sqrt{2bu^2}r\right)$.
Hence, we see that the superconducting order parameter asymptotically behaves as
\begin{equation}
    \psi(r) \sim u + c_{\psi} K_0\left(\sqrt{2bu^2}r\right).
\end{equation}
For the in-plane gauge field $\vec{\alpha}_{r\theta}=\alpha(r)\vec{e}_{\theta}$, the linearized field equation becomes a modified Bessel equation of first order
\begin{equation}
    r^2\frac{\textup{d}^2\alpha}{\textup{d} r^2} + r\frac{\textup{d} \alpha}{\textup{d} r} - \left(q^2u^2r^2 + 1\right)\alpha = 0,
\end{equation}
which has solution $\alpha(r) = c_{A} K_1(qur)$.
Therefore, the in-plane gauge field has the asymptotic behaviour
\begin{equation}
    \vec{A}_{r\theta}(r) \sim c_{A} K_1(qur) \vec{e}_{\theta}.
\end{equation}
This is identical to the asymptotic behaviour of the superconducting order parameter and in-plane gauge field in the Abelian Higgs model \cite{Speight_1997,Manton_Speight_2003,Bettencourt_1995,Fujikura_2023}.

\begin{figure}[t]
    \centering
    \includegraphics[width=\linewidth]{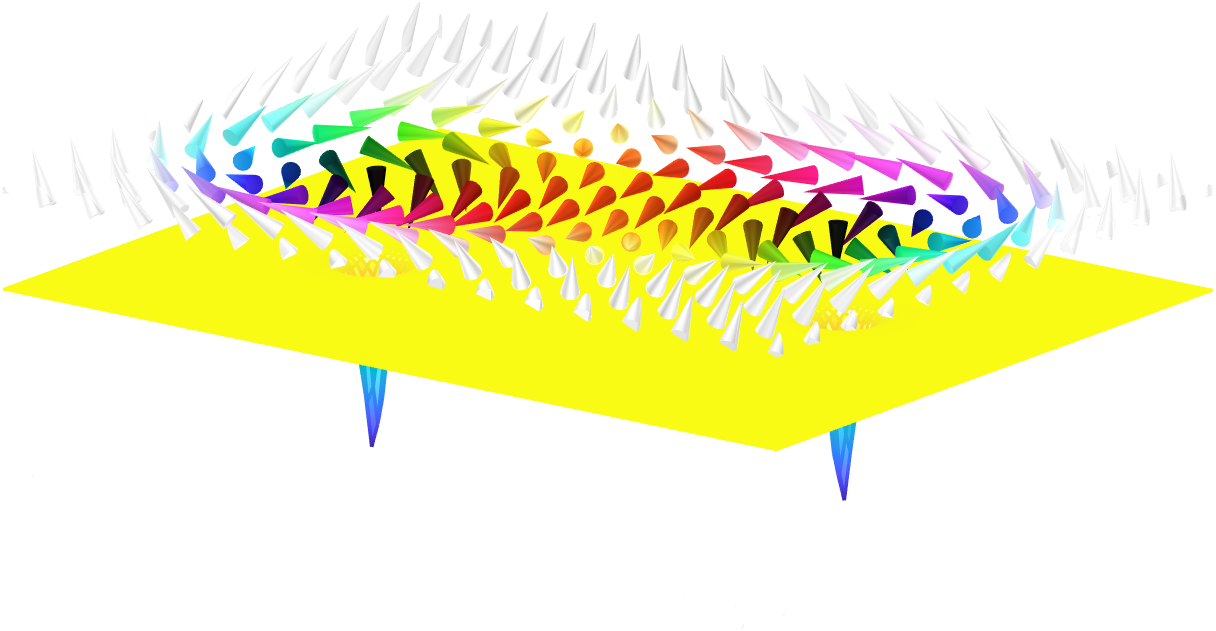}
    \caption{Depiction of the bulk magnetic spin texture and superconducting order parameter density for the SVP-SVP bound state. The Bloch configuration for the magnetization is the skyrmion type that yields the lowest energy SVP, in comparison to N\'eel and anti-skyrmions.} 
\label{fig: SVP bound state}
\end{figure}

For the magnetization we have various skyrmion ans\"atze we could employ, e.g. Bloch, N\'eel or antiskyrmion.
However, we observe that the Bloch ansatz produces the lowest energy skyrmion numerically, and is displayed in Fig.~\ref{fig: SVP bound state}.
Therefore, we consider Bloch perturbations of the form $\vec{n}=f(r)\vec{e}_{\theta}$.
This yields a coupled system of ODEs for the magnetization $f(r)$ and out-of-plane gauge field $\alpha_z(r)$,
\begin{align}
    \frac{\textup{d}^2f}{\textup{d} r^2} + \frac{1}{r}\frac{\textup{d} f}{\textup{d} r} - \frac{1}{r^2}f - \frac{\textup{d} \alpha_z}{\textup{d} r} = \, & 0, \\
    \frac{\textup{d}^2\alpha_z}{\textup{d} r^2} + \frac{1}{r}\frac{\textup{d} \alpha_z}{\textup{d} r} - q^2u^2\alpha_z + \frac{\textup{d} f}{\textup{d} r} + \frac{f}{r} = \, & 0.
\end{align}
This coupled system can be solved analytically.
The general asymptotic behaviour of the out-of-plane gauge field is found to be given by
\begin{equation}
    A_z(r) \sim -\frac{c_{m}}{\sqrt{q^2u^2 - 1}}K_0\left( \sqrt{q^2u^2 - 1}\,r \right),
\end{equation}
and the magnetization asymptotically behaves as 
\begin{equation}
    \vec{m}(r) \sim \vec{m}_0 + \frac{c_{m}}{q^2u^2 - 1} K_1\left( \sqrt{q^2u^2 - 1}\,r \right) \vec{e}_{\theta}.
\end{equation}

In general, the long-range interaction of solitons is governed by the decay lengths of the linearized modes about the uniform ground state, rather than by the coherence lengths associated with amplitude relaxation of the order parameters. 
When the magnetization length is fixed, longitudinal fluctuations are frozen out and there is no independent magnetic coherence length.
The only remaining magnetic length scale is the decay length governing transverse distortions of the magnetization.

From the asymptotic forms, we are able to deduce the relevant length scales of the fields.
The superconductor and magnetization decay lengths are $\xi_s$ and $ l_m $, respectively.
We have two magnetic penetration depths, one corresponding to the perpendicular magnetic field, $\lambda$, and one to the in-plane magnetic field, $\lambda_{\parallel}$.
The length scales are determined as
\begin{equation}
\label{eq: Length scales}
    \xi_s = \frac{1}{\sqrt{2bu^2}}, \quad
    \lambda=\frac{1}{qu}, \quad \lambda_{\parallel}= l_m =\frac{1}{\sqrt{q^2u^2-1}}.
\end{equation}
We note that the magnetization decay length $ l_m $ is real for $qu>1$.
In terms of the length scales \eqref{eq: Length scales}, the asymptotic forms can be summarized as
\begin{subequations}
\label{eq: Asymptotic forms}
    \begin{align}
        \label{eq: Asymptotic OP}
        \phi(r) = \, & c_{\psi} K_0\left(\frac{r}{\xi_s}\right), \\
        \label{eq: Asymptotic gauge}
        \vec{\alpha}(r) = \, & c_{A} K_1\left(\frac{r}{\lambda}\right) \vec{e}_{\theta} - c_m  l_m  K_0\left( \frac{r}{ l_m } \right) \vec{e}_z, \\
        \label{eq: Asymptotic magnet}
        \vec{n}(r) = \, &  c_m  l_m ^2 K_1\left( \frac{r}{ l_m } \right)\vec{e}_{\theta}.
    \end{align}
\end{subequations}


\subsection{Long-range interaction energy of composite states}
\label{subsec: Long-range interaction energy of composite states}

\begin{figure}[t]
    \centering
    \includegraphics[width=\linewidth]{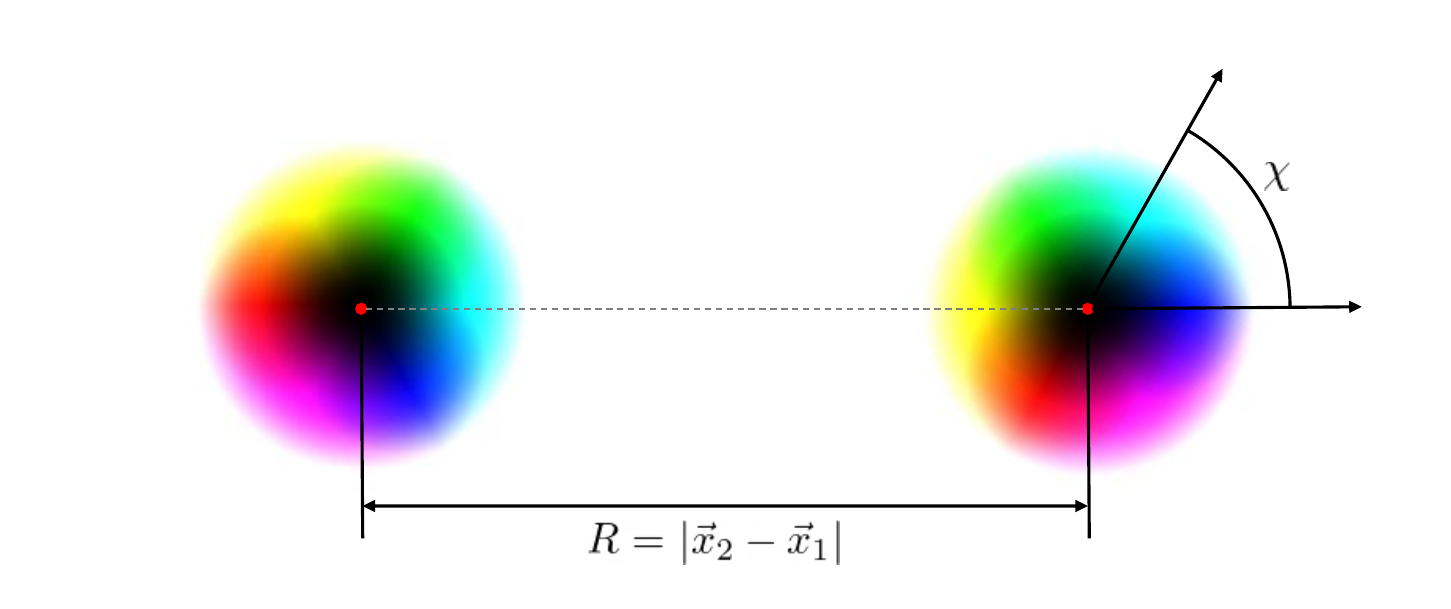}
    \caption{Setup for computing the long-range interaction between two well-separated composite skyrmion-vortex pairs. One pair is located at $\vec{x}_1$ and the other pair at $\vec{x}_2$. The distance between the pair is $R=|\vec{x}_2-\vec{x}_1|$, with the red dots indicating the zeroes of the order parameter $\psi$, i.e. the vortex locations. There is a relative phase orientation between the two skyrmions, characterized by the rotation angle $\chi\in[0,2\pi)$ that acts on the in-plane magnetization $(m_x,m_y)$-components, via a rotation matrix $R_z(\chi)\in\textup{SO}(2)$.} 
\label{fig: SVP-SVP}
\end{figure}

In order to understand the long-range interactions of well-separated SVPs, we must construct a linearized field theory such that its solutions are identical to the asymptotic forms \eqref{eq: Asymptotic forms} of the SVP determined in the previous section.
To do this, we introduce an external source $\mathcal{F}_{\textup{source}} = -\rho\phi - j_i \alpha_i - \sigma_i n_i$ and consider the following energy
\begin{align}
\label{eq: Sourced energy}
    \mathcal{F} = \, & \mathcal{F}_{\textup{lin}} +\mathcal{F}_{\textup{source}} \nonumber \\
    = \, & \frac{1}{2} \phi \left( -\Delta + \frac{1}{\xi_s^2} \right) \phi + \frac{1}{2}\vec{\alpha} \cdot \left( -\Delta + \frac{1}{\lambda^2} \right) \vec{\alpha}  \nonumber \\
    \, & + \frac{1}{2}\vec{n}\cdot \left(-\Delta\right) \vec{n} - \vec{n}\cdot \left( \vec{\nabla} \times \vec{\alpha} \right) -\rho\phi - \vec{j}\cdot\vec{\alpha} - \vec{\sigma} \cdot \vec{n}.
\end{align}
Upon variation of this new energy \eqref{eq: Sourced energy} with respect to the field perturbations $\{\phi,\vec{\alpha},\vec{n}\}$, we obtain a modified system of coupled ODEs,
\begin{subequations}
    \begin{align}
        \label{eq: Linearized OP eqn}
        \left( -\Delta + \frac{1}{\xi_s^2} \right) \phi = \, & \rho, \\
        \label{eq: Linearized gauge eqn}
        \left( -\Delta + \frac{1}{\lambda^2} \right) \vec{\alpha} = \, & \vec{j} + \vec{\nabla} \times \vec{n} - \vec{\nabla}\left( \vec{\nabla} \cdot \vec{\alpha} \right), \\
        \label{eq: Linearized magnet eqn}
        -\Delta\vec{n} = \, & \vec{\sigma} + \vec{\nabla} \times \vec{\alpha}.
    \end{align}
\end{subequations}
We need to solve this system for the sources $\{\rho,\vec{j},\vec{\sigma}\}$ using the previously determined asymptotic forms \eqref{eq: Asymptotic forms}.

The Green's function for the static Klein-Gordon equation in 2D is the Bessel function $K_0$, that is, it satisfies
\begin{equation}
    \left( -\Delta + \lambda^2 \right) K_0(\lambda r) = 2\pi\delta(r).
\end{equation}
Substituting the asymptotic form of the superconducting order parameter \eqref{eq: Asymptotic OP} into the modified field equation \eqref{eq: Linearized OP eqn} yields
\begin{equation}
    \rho(r) = \left( -\Delta + \frac{1}{\xi_s^2} \right)c_{\psi} K_0\left(\frac{r}{\xi_s}\right) = c_{\psi} 2\pi \delta(r).
\end{equation}
A similar approach allows us to determine the other two additional sources,
\begin{align}
    \label{eq: External source j}
    \vec{j}(r) = \, & -2\pi c_A \lambda \left[\vec{e}_z \times \vec{\nabla}\delta(r)\right] -2\pi c_m  l_m \delta(r)\vec{e}_z, \\
    \label{eq: External source sigma}
    \vec{\sigma}(r) = \, & \frac{c_A}{\lambda} K_0\left(\frac{r}{\lambda}\right)\vec{e}_z.
\end{align}
The last contribution in \eqref{eq: External source j} comes from the Zeeman interaction.

With the asymptotic forms and additional sources now in-hand, we can now compute the asymptotic interaction energy of well-separated SVPs.
Let us consider a SVP at $\vec{x}_1$ and label that pair as SVP$^{(1)}$, and another pair SVP$^{(2)}$ at $\vec{x}_2$.
Further, we allow a relative $\textup{SO}(2)_{\textup{iso}}$ iso-rotation of the separated skyrmions.
This is depicted in Fig.~\ref{fig: SVP-SVP}.
We parameterize this by a rotation angle $\chi\in[0,2\pi)$ that acts on the in-plane magnetization ($n_r, n_\theta$) components of, say, SVP$^{(1)}$.
The corresponding magnetization at SVP$^{(1)}$ is thus given by
\begin{equation}
    \vec{n}^{(1)}(\chi,r_1) = c_m  l_m ^2 K_1 \left( \frac{r_1}{ l_m } \right) \left[ -\sin(\chi) \vec{e}_r + \cos(\chi)\vec{e}_{\theta} \right],
\end{equation}
where $r_1=|\vec{x}-\vec{x}_1|$.
We can express this in terms of the rotation matrix
\begin{equation}
    R_z(\chi)=
    \begin{pmatrix}
        \cos\chi & -\sin\chi \\
        \sin\chi & \cos\chi
    \end{pmatrix}
    \in\textup{SO}(2)
\end{equation}
as $\vec{n}^{(1)}(\chi,r_1)=R_z(\chi)\vec{n}(r_1)$.
The Zeeman interaction breaks the $\textup{SO}(2)$ isospin symmetry of this model, due to the presence of vortices.
So, in order to keep the Zeeman interaction pointwise invariant under the $\textup{SO}(2)$ action, we require
\begin{equation}
    \vec{B}^{(1)}(\chi,r_1)=R_z(\chi)\vec{B}(r_1).
\end{equation}
A convenient way to realize this on the gauge field $\vec{\alpha}(r_1)$ is simply to co-rotate $\vec{\alpha}(r_1)$ by the same $\textup{SO}(2)$ rotation.
That is, we define $\vec{\alpha}^{(1)}(\chi,r_1)=R_z(\chi)\vec{\alpha}(r_1)$ such that
\begin{equation}
    \vec{\nabla}\times\vec{\alpha}^{(1)}(\chi,r_1) = R_z(\chi)(\vec{\nabla}\times\vec{\alpha})=\vec{B}^{(1)}(\chi,r_1).
\end{equation}
Explicitly, the gauge field at SVP$^{(1)}$ is given by
\begin{align}
    \vec{\alpha}^{(1)}(\chi,r_1) = \, & c_{A} K_1\left( \frac{r_1}{\lambda} \right) \left[ -\sin(\chi) \vec{e}_r + \cos(\chi)\vec{e}_{\theta} \right] \nonumber \\
    \, & - c_m  l_m  K_0\left( \frac{r}{ l_m } \right) \vec{e}_z.
\end{align}
Furthermore, we must also co-rotate the external current $\vec{j}$ such that the field equation \eqref{eq: Linearized gauge eqn} remains invariant.
It is straightforward to see that the relevant co-rotation is simply $\vec{j}^{(1)}(\chi,r_1)=R_z(\chi)\vec{j}(r_1)$.

The interaction energy between well-separated sources comes from the cross-terms in the linearization.
That is,
\begin{align}
    E_{\textup{int}} = \, & -\int_{\mathbb{R}^2}\textup{d}^2\vec{x}\, \left\{\rho^{(1)} \phi^{(2)} +\vec{j}^{(1)}\cdot\vec{\alpha}^{(2)} + \vec{n}^{(1)}\cdot\vec{\sigma}^{(2)} \right. \nonumber \\
    \, & \left. + \vec{n}^{(1)}\cdot (\vec{\nabla\times}\vec{\alpha})^{(2)} \right\}.
\end{align}
After a bit of work we arrive at the interaction energy in terms of SVP separation $R=|\vec{x}_2-\vec{x}_1|$ and relative skyrmion orientation $\chi$,
\begin{align}
    \, & E_{\textup{int}}(R,\chi) = 2\pi \left\{c_A ^2 K_0\left( \frac{R}{\lambda} \right) - c_{\psi}^2 K_0\left(\frac{R}{\xi_s}\right) \right\} \nonumber \\
    \, & - 2\pi c_m^2  l_m ^2 K_1\left( \frac{R}{ l_m } \right) + \pi^2 c_m^2  l_m ^4 K_1\left(\frac{R}{ l_m }\right) \cos(\chi).
\end{align}
The first two contributions are from the usual Ginzburg--Landau vortex-vortex interaction, the third term arises from the Zeeman interaction, and the final term is the skyrmion-skyrmion interaction.

Let us first consider the impact the skyrmion iso-rotation angle $\chi\in[0,2\pi)$ has on the interaction energy.
We have that
\begin{subequations}
    \begin{align}
        \frac{\partial E_{\textup{int}}}{\partial \chi} = \, & -\pi^2 c_m^2  l_m ^4 K_1\left(\frac{R}{ l_m }\right) \sin(\chi), \\
        \frac{\partial^2 E_{\textup{int}}}{\partial \chi^2} = \, & -\pi^2 c_m^2  l_m ^4 K_1\left(\frac{R}{ l_m }\right) \cos(\chi).
    \end{align}
\end{subequations}
Clearly, the interaction energy is extremized for the choice $\chi=k\pi$ with $k\in\{0,1\}$.
The Hessian $\partial^2 E_{\textup{int}}/\partial \chi^2$ is positive definite if $k=1$, i.e. $\chi=\pi$ minimizes the interaction energy.
That is, the interaction energy is minimized if one skyrmion is iso-rotated by $180^\circ$ relative to the other.
This is observed numerically, with a SVP-SVP bound state shown in Fig.~\ref{fig: N=2,Q=2 SC}.

Next, we consider the dominant contribution in the interaction energy.
To leading order, the large $R$ behaviour of the Bessel functions are
\begin{equation}
    K_{0,1}(\mu R) = e^{-\mu R}\left[ \sqrt{\frac{\pi}{2R\mu}} + O(R^{-3/2}) \right]
\end{equation}
Then, the interaction energy to leading order can be expressed as
\begin{align}
\label{eq: Leading order interaction energy}
    E_{\textup{int}}(R,\chi) \approx \, & \sqrt{\frac{\pi^3}{2R}} \left\{ c_m^2 l_m ^2 \sqrt{ l_m } \left[ \pi  l_m ^2\cos(\chi) - 2 \right]e^{-\frac{R}{ l_m }} \right. \nonumber \\
    & \left.+ 2c_A^2 \sqrt{\lambda} e^{-\frac{R}{\lambda}} - 2c_\psi^2 \sqrt{\xi_s} e^{-\frac{R}{\xi_s}} \right\}.
\end{align}
Let us consider the interaction energy in the attractive channel (that is, the energy-minimizing isorotation $\chi=\pi$ orientation).
Then, the Zeeman-induced mixed mode contributes a strictly attractive Yukawa tail.
So the long-range sign of the SVP-SVP interaction is determined purely by which decay length is largest: if $l_m$ (or $\xi_s)$ exceeds $\lambda$, the interaction is attractive for large separation $R$.
However, for $qu>1$, it is always true $l_m>\lambda$.
Therefore, the interaction energy at long-range is attractive.

\begin{figure*}[t]
    \centering
    \begin{subfigure}[b]{0.8\textwidth}
        \includegraphics[width=\textwidth]{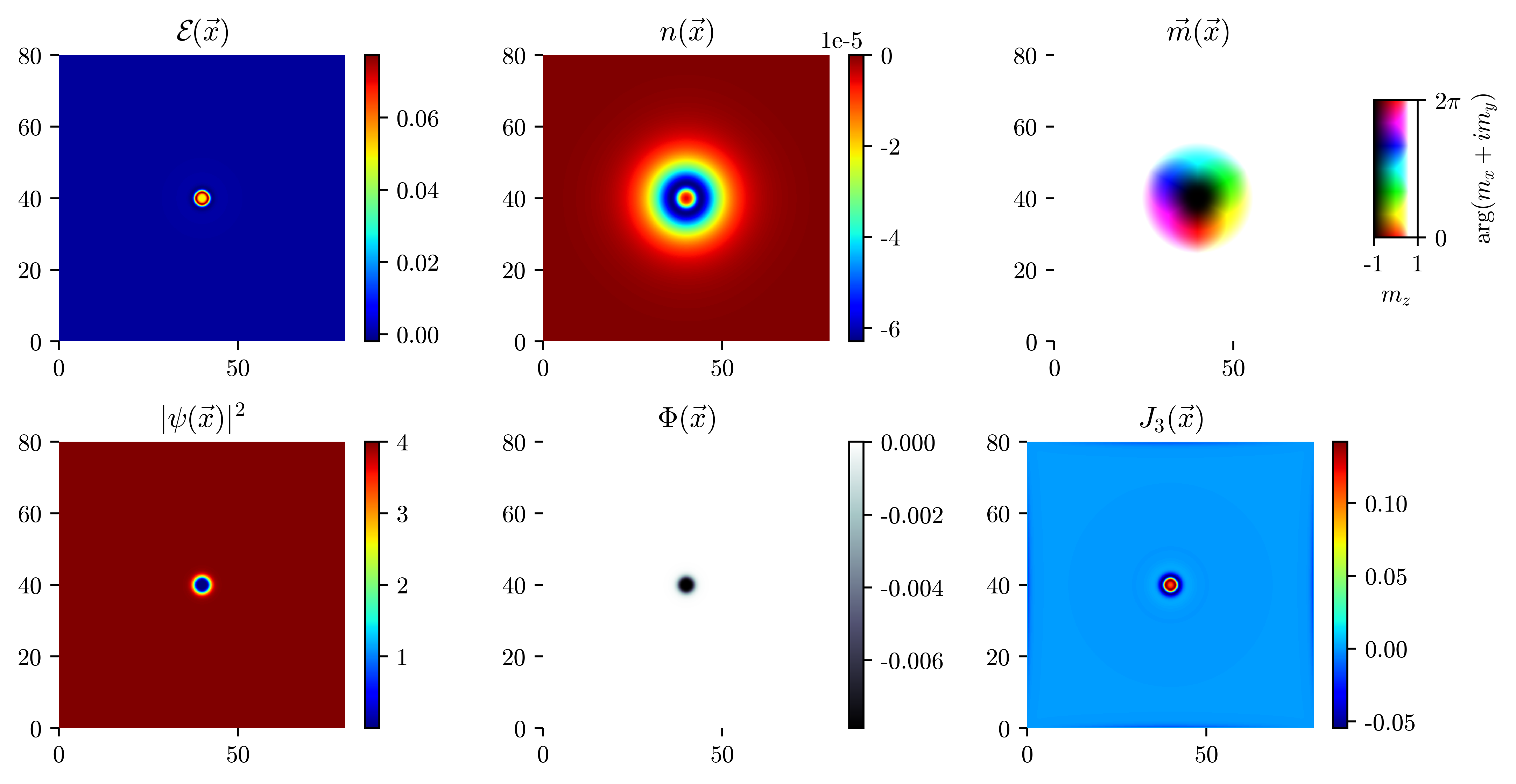}
        \caption{}
    \end{subfigure}
    \\
    \begin{subfigure}[b]{0.8\textwidth}
        \includegraphics[width=\textwidth]{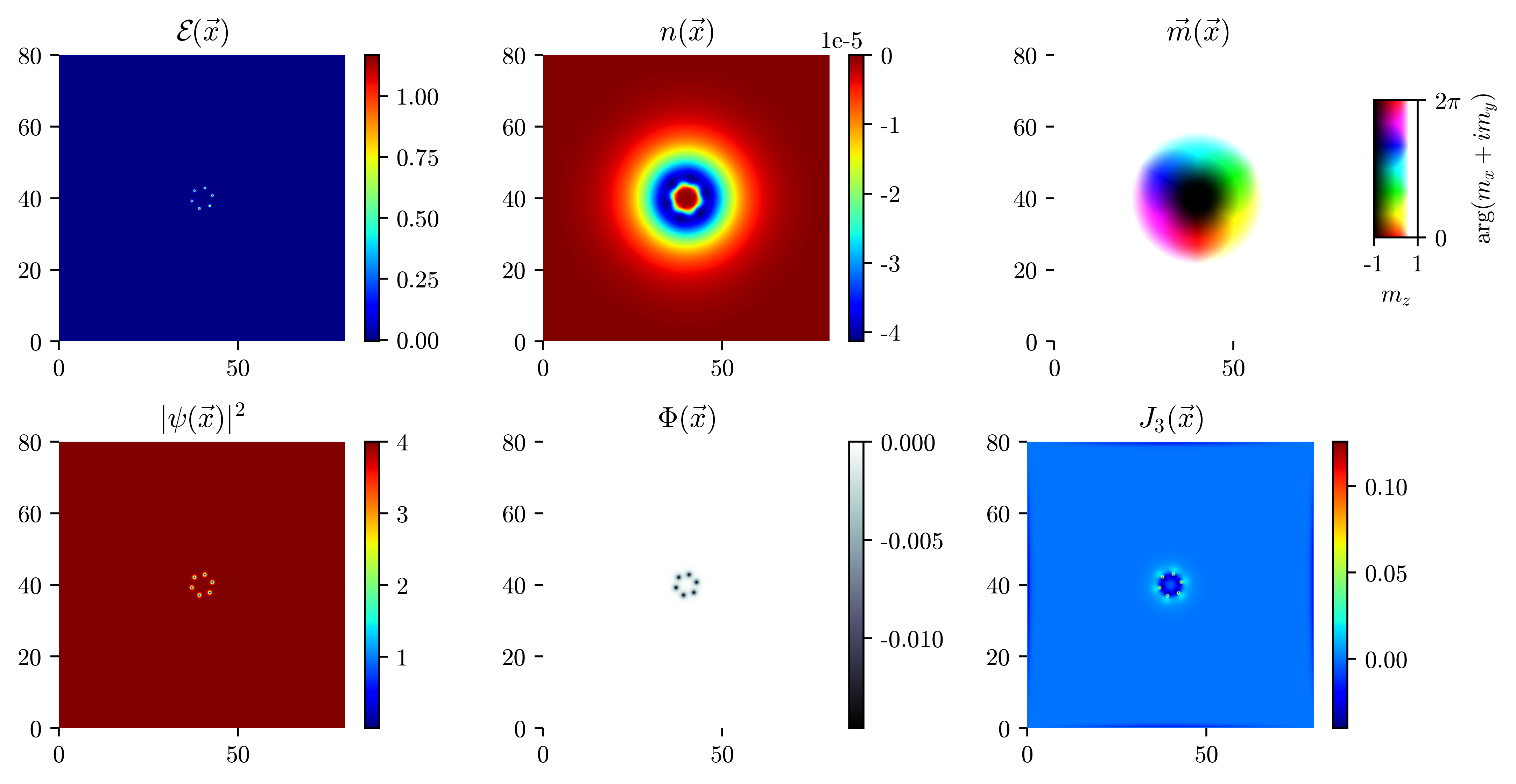}
        \caption{}
    \end{subfigure}
    \caption{Comparison of composite skyrmion-mulitvortex configurations in the different regimes. Shown here is a 6-vortex configuration coinciding with a single Block skyrmion. The parameters for the magnetization are fixed with $\beta=-\alpha=1$, such that the ground state magnetization has value $m_0^2=1$. For the superconducting order parameter, we fix the ground state value $u^2=-a/b=4$ and consider two cases. The first is (a) the type-I regime (corresponding to $b<\tfrac{1}{2}q^2$) where vortex cores coincide to form one big axially symmetric state, with $\lambda<\xi_s$. The second is (b) the bound state regime (corresponding to $b>\tfrac{1}{2}q^2$) where vortex clustering occurs, with $\xi_s<\lambda< l_m $.}
    \label{fig: Type 1.5 regime}
\end{figure*}

Consider now the two terms contributing to the vortex-vortex interaction: the scalar attraction and magnetic repulsion.
These are proportional to
\begin{equation}
    E_{VV}(R)=c_A^2 \sqrt{\lambda} e^{-\frac{R}{\lambda}} - c_\psi^2 \sqrt{\xi_s} e^{-\frac{R}{\xi_s}}.
\end{equation}
The first term originates from the gauge field, it repels vortices due to circulating currents.
The second is an attractive force coming from core-core interactions.
When the scalar attraction from the core-core interaction dominates, the force $-E'_{VV}(R)$ between vortices is attractive and the vortex cores (zeroes of the order parameter $\psi$) coincide.
This occurs when $\lambda < \xi_s$.
On the other hand, when $\lambda > \xi_s$, the magnetic repulsion dominates and force between vortices is repulsive.
To have bound states with intermediate-range repulsion and long-range attraction, we require the magnetic penetration depth to be between the two decay lengths, $\xi_s<\lambda< l_m $, where a partial analogy can be driven to type-1.5 superconductivity \cite{Babaev_Speight_2005}.
As we have already established, for $qu>1$ it is always true that the magnetization decay length is larger than the magnetic penetration depth, $\lambda< l_m $.
Therefore, for vortex clustering we only need $\lambda>\xi_s$, which amounts to choosing $b>\tfrac{1}{2}q^2$.
This is illustrated in Fig.~\ref{fig: Type 1.5 regime}.

One can compare the bound states between skyrmion-vortex pairs in the ferromagnetic superconductor model to type-1.5 behavior of composite vortices in multicomponent superconductors \cite{Babaev_2002,Babaev_Speight_2005,Babaev_Carlstrom_2010,Babaev_2011,Babaev_2017}.
In both cases, the interaction between topological defects is governed by the competition between multiple decay lengths associated with distinct collective modes.
In multiband superconductors, this competition arises from multiple scalar amplitude modes of different condensates and the magnetic penetration depth, leading to long-range attraction and short-range repulsion \cite{Babaev_2013}.
In the present model, also a multi-scale mechanism operates: the superconducting condensate, gauge field, and magnetization-gauge mixed mode each generate Yukawa-type interactions with different characteristic lengths.
When the longest decay length corresponds to an attractive channel, most notably the mixed magnetization-gauge mode at the energetically preferred isorotation, the interaction becomes attractive at long range, while repulsive contributions dominate at shorter distances.
This interplay naturally produces non-monotonic forces and stable bound states.

\section{Conclusion}
\label{sec: Conclusion}

We have presented a self-consistent field-theoretical study of composite skyrmion-vortex pairs in ferromagnetic superconductors, showing that they form stable bound states through the interplay of superconducting and magnetic order parameters.
By deriving the asymptotic forms of the coupled fields, we established the existence of multiple characteristic length scales that govern the long-range interaction of SVPs.
This revealed a competition between short-range repulsion and long-range attraction, leading to a non-monotonic interaction potential between well-separated SVPs.
This provides a mechanism for the formation of bound states and clustering of SVPs in certain parameter regimes.

Furthermore, we find that the interaction depends on the relative internal orientation of the magnetic skyrmion spin textures, with a $\pi$ relative isorotation corresponding to the energetically preferable orientation, known as the attractive channel orientation.
This orientation dependence is analogous to the behavior of skyrmions in the baby Skyrme model.
In the present case, this orientation dependence arises within a composite excitation and is mediated by the coupling between magnetic and superconducting degrees of freedom.

Overall, our results establish a self-consistent theoretical framework for understanding interactions between composite skyrmion-vortex excitations in bulk ferromagnetic superconductors.
It highlights the importance of treating superconducting and magnetic degrees of freedom self-consistently when analyzing their energetics and collective behavior.


\section*{Acknowledgments}

PL acknowledges funding from the Olle Engkvists Stiftelse through the grant 226-0103.
CR acknowledges funding from the Edge Hill Research Investment Fund.
EB was supported by the Swedish Research Council Grants 2022-04763, a project grant from Knut och Alice Wallenbergs Stiftelse, and partially by the Wallenberg Initiative Materials Science for Sustainability (WISE) funded by the Knut and Alice Wallenberg Foundation.


\section*{Data availability}

The code used to produce the data in this text can be found at the publicly available github repository \href{https://github.com/Paulnleask/cuSuperFerro}{https://github.com/Paulnleask/cuSuperFerro}.


\bibliography{main.bib}

\end{document}